\documentclass{article}

\usepackage{arxiv}

\usepackage[utf8]{inputenc} 
\usepackage[T1]{fontenc}    
\usepackage{hyperref}       
\usepackage{url}            
\usepackage{booktabs}       
\usepackage{amsfonts}       
\usepackage{nicefrac}       
\usepackage{microtype}      
\usepackage{lipsum}

\usepackage{amsmath}
\usepackage{siunitx}
\usepackage{longtable}
\usepackage{array}
\usepackage{multirow}
\usepackage{wrapfig}
\usepackage{float}
\usepackage{colortbl}
\usepackage{pdflscape}
\usepackage{tabu}
\usepackage{threeparttable}
\usepackage{threeparttablex}
\usepackage[normalem]{ulem}
\usepackage{makecell}
\usepackage{xcolor}
\usepackage[round]{natbib}
\usepackage{enumerate}
\usepackage[shortlabels]{enumitem}

\usepackage{pgfplots}
\usepackage{pdflscape}
\title{Optimal sampling for design-based estimators of regression models}

\author{
  Tong Chen \\
  Department of Statistics\\
  University of Auckland\\
  Email:tche929@aucklanduni.ac.nz\\
   \And
 Thomas Lumley \\
  Department of Statistics\\
  University of Auckland\\
  Email:t.lumley@auckland.ac.nz\\
}

\begin{document}
\maketitle

\begin{abstract}
Two-phase designs measure variables of interest on a subcohort where the outcome and covariates are readily available or cheap to collect on all individuals in the cohort. Given limited resource availability, it is of interest to find an optimal design that includes more informative individuals in the final sample. We explore the optimal designs and efficiencies for analysis by design-based estimators. Generalized raking is an efficient design-based estimator that improves on the inverse-probability weighted (IPW) estimator by adjusting weights based on the auxiliary information. We derive a closed-form solution of the optimal design for estimating regression coefficients from generalized raking estimators. We compare it with the optimal design for analysis via the IPW estimator and other two-phase designs in measurement-error settings. We consider general two-phase designs where the outcome variable and variables of interest can be continuous or discrete. Our results show that the optimal designs for analysis by the two design-based estimators can be very different.  The optimal design for IPW estimation is optimal for analysis via the IPW estimator and typically gives near-optimal efficiency for generalized raking, though we show there is potential improvement in some settings.
\end{abstract}


\section{Introduction}
\label{sec1}
In modern public health studies, routinely collected large databases, such as electronic health records (EHR), are increasingly used to study research questions of interest. However, variables within these databases can be error-prone. Without validation, directly analyzing EHR data may lead to invalid statistical inference. For these large databases, it will be prohibitively expensive to validate the variables of interest for the entire cohort. A cost-effective strategy is to use two-phase sampling \citep{neyman1938contribution}. At phase one, the outcome variable and several covariates (e.g., age, gender, and ethnicity) are collected or available for every individual in the cohort. The EHR databases can be used as the phase-1 sample. At phase two, variables of interest, such as biomarkers, are collected and validated for individuals selected in the phase-2 subsample. 

Design-based methods produce robust estimations for fitting regression models to two-phase stratified sampling. Directly solving inverse-probability weighted (IPW) or Horvitz-Thompson \citep{horvitz1952generalization} type of likelihood functions leads to the IPW estimator. For analyzing EHR databases, the IPW estimator is not efficient as most of the information in the phase-1 sample has been ignored. Generalized raking \citep{deville1992calibration,robins1994estimation} is more efficient than the IPW estimator as it incorporates the whole cohort information in the analysis. The efficiency gains are achieved by adjusting the sampling probabilities of the IPW estimator based on auxiliary information available for the whole cohort. Generalized raking estimators are closely connected with the augmented inverse-probability weighted (AIPW) estimators of \cite{robins1994estimation} \citep{lumley2011connections}. Typically, generalized raking estimators are combined with imputations to solve problems of fitting regression models in two-phase designs. See \cite{breslow2009improved,breslow2009using} for single imputation, and \cite{oh2020improved} and \cite{han2019combining} for multiple imputation. 

An alternative estimation method is based on semi-parametric maximum likelihood \citep{scott1997fitting, tao2017efficient}, which can be more efficient than design-based estimators if the model is correctly specified. \cite{han2019combining} showed generalized raking estimators can be more efficient than the semi-parametric maximum likelihood estimator even under mild model misspecification.

The optimal sampling theory closely connected with design-based estimators has been studied in some previous literature. \cite{reilly1995mean} derived a closed-form expression of the optimal phase-two sampling probabilities for the mean-score estimator. \cite{mcisaac2015adaptive} and \cite{han2020two} extended the work using a multiwave sampling framework for binary and survival outcomes respectively. The optimal design of estimating regression coefficients from the IPW estimator is Neyman allocation \citep{neyman1934two} applied to influence functions. For binary data and stratified sampling, the optimal design for analysis by the mean-score estimator is asymptotically identical to those for analysis by the IPW estimator \citep{mcisaac2014response, chen2020multi}. \cite{mcisaac2014response} suggested the optimal design for analysis by the AIPW estimator can be derived numerically. If variables are all discrete, it is asymptotically equivalent to the optimal designs for analysis by the mean score and the IPW estimator. However, a closed-form expression of the optimal allocation for analysis via generalized raking is not known.

In this article, we derive a closed-form expression of the optimal allocation for analysis via generalized raking estimators. We then compare it with the optimal allocation for analysis by the IPW estimator and other commonly used sampling designs. Furthermore, sampling probabilities are assumed to be known in a two-phase design, so that we do not consider doubly robustness \citep{bang2005doubly}. The rest of this article is organized as follows. Notations are defined in Section \ref{sec2}. The IPW estimator and generalized raking estimators are introduced in Section \ref{dbe}. In Section \ref{optd}, we derive the optimal designs for analysis by the two design-based estimators. Results of simulation studies are reported in Section \ref{sims}. In Section \ref{dataeg}, we further compare the proposed optimal designs with other sampling strategies using the National Wilms' Tumor Study (NWTS) \citep{d1989treatment,green1998comparison} dataset example. Remarks are made in Section \ref{final}. Code of numerical studies is available from \url{https://github.com/T0ngChen/Opt_sampling_design_based}. An interactive \texttt{Shiny} \citep{shinypackage} app, which compares the optimal designs for analysis by the two classes of design-based estimators, is available from \url{https://tchen.shinyapps.io/raking}.

\section{Notation} \label{sec2}
Suppose we want to select $n$ observations with stratified random sampling from a cohort of size $N$ over $K$ strata. Let $N_i$ and $n_i$ be the stratum size and phase-2 sample size for stratum $i$ respectively. Let $Y$ denote an outcome variable, $Z$ denote covariates collected at phase 1, and $A$ denote auxiliary variables. Variables $Y$, $Z$, and $A$ are available for every individual in the phase-1 sample. Let $X$ represent variables of interest which are only available for individuals selected in the phase-2 sample. Let $R$ denote an indicator variable. If $R_i = 1$, subject $i$ is selected in the phase-2 sample, otherwise $R_i = 0$. We assume the missingness of $X$ only depends on phase-1 data, $P(R|X,Y, A, Z) = P(R|Y,A,Z)$, so variables of interest $X$ are missing at random \citep{rubin1976inference}. The phase-2 inclusion probability of individual $i$ is $E(R_i|Z_i,A_i,Y_i) = \pi_i$.

We describe $P(X|A,Z;\alpha)$ as the imputation model and $P(Y|X,Z;\beta)$ as the outcome model of interest, where $\alpha$ and $\beta$ are regression coefficients in the imputation model and the outcome model respectively.  Specifically, let $\beta_1$ be the regression coefficient of $X$ in the outcome model. Our target is to minimize the variance of $\hat{\beta}_1$ by optimizing the design for design-based estimators. 

\section{Design-based estimators}\label{dbe}
\subsection{Inverse-probability weighted estimator}

The IPW \citep{horvitz1952generalization} estimator can be derived by weighting each observation by the inverse of its sampling probability $\pi_i$. The IPW estimator for estimating regression coefficients $\beta$ can be obtained by solving weighted score function
\begin{equation}
    \sum_{i=1}^{N} \frac{R_i}{\pi_i} \log P\left(Y_{i} \mid X_{i}, Z_{i} ; \beta\right) / \partial \beta=0. \label{w_likeli}
\end{equation}
The sampling probability $\pi_i$ in Equation (\ref{w_likeli}) should be bounded away from zero so that every study subject should have a positive probability of being sampled at phase two. The IPW estimator is appealing because of its simplicity and robustness. However, it is not efficient as it ignores the phase-1 information.

\subsection{Generalized raking estimator}\label{gr-procedure}
Generalized raking estimators improve on the IPW estimator by adjusting sampling weights based on the auxiliary information. Suppose population totals of a vector of auxiliary variables $S$ are known in advance, and we want to estimate the population total $T_X = \sum_{i = 1}^N X_i$. Generalized raking estimators are defined as $T_{xr} = \sum_{i=1}^N R_iw_iX_i$, where $w_i$ are calibrated weights, and they only depend on auxiliary variables $S$. The target is to minimize the total weight change $\sum_{i=1}^N R_i d(w_i,1/\pi_i)$ with calibration constraints
$$
\sum_{i=1}^N R_iw_iS_i = \sum_{i=1}^N S_i,
$$

under a prespecified distance function $d(a,b)$. The optimization problem can be solved by Lagrange multipliers. \cite{deville1992calibration} provided a few example distance functions and showed all generalized raking estimators are asymptotically equivalent. Typically, distance function $d(a,b) = (a-b)^2/2b$ will lead to the \textit{generalized regression estimator} (GREG).

We want to use generalized raking procedures to estimate the regression parameter of interest $\beta$. According to the generalized raking procedures described above, population totals or means of auxiliary variables should be good approximations of $\beta$ and known in advance. An asymptotically linear estimator $\hat{\beta}$ satisfies
\begin{align}
    \sqrt{N}(\hat{\beta}-\beta)=\frac{1}{\sqrt{N}} \sum_{i=1}^{N} \mathbf{h}_i(\beta) +o_{p}(1), \label{inf}
\end{align}
where $\mathbf{h}_i(\beta)$ is the influnce function for $i^{th}$ observation. \cite{breslow2008az} derived a weighted version of Equation (\ref{inf}) for the IPW estimator. As the asymptotically linear estimator $\hat{\beta}$ can be approximated by the mean of influence functions, good choices of auxiliary variables should be strongly correlated with influence functions. A generalized raking estimator will be asymptotically efficient among all design-based estimators if auxiliary variables are $E(\mathbf{h}_i(\beta)|Y,Z,A)$ \citep{lumley2011connections}.

Influence functions depend on unknown parameters $\beta$, so optimal auxiliary variables are typically unavailable. \cite{kulich2004improving} proposed a ``plug-in'' method to approximate the conditional expectation $E(\mathbf{h}_i(\beta)|Y,Z,A)$ where missing $X$ are imputed using data from the phase-2 sample, and $\mathbf{h}_i(\hat{\beta})$ are estimated using the imputed $X$ and phase-1 data \citep{breslow2009using, rivera2016using}. In this article, we followed the same procedures to get generalized raking estimations in the final analysis. The procedures are summarized as follows:

\begin{enumerate}
    \item Fit a weighted regression model (imputation model) using phase-2 data, and impute $X$ for all individuals. 
    \item Fit the outcome model with phase-1 data and imputed $X$, and then estimate influence functions $\mathbf{h}_i(\hat{\beta})$. For linear regression, $\mathbf{h}_i(\hat{\beta})$ can be obtained using \textit{dfbeta} in \texttt{R} \citep{Rcore}. For other regression models, we can get $\mathbf{h}_i(\hat{\beta})$ from the score or jackknife estimators. 
    \item Estimate $\beta$ using generalized raking with adjusted weights. We use the distance function $d(a,b) = a \log(a/b)-a+b$, which makes the calibrated weights $w_i$ non-negative. 
\end{enumerate}

The \texttt{survey} package \citep{lumley2011complex} in \texttt{R} \citep{Rcore} is used to fit the weighted regression in step 1 and obtain generalized raking estimations in step 3. When fitting the outcome model in step 2, note that it is crucial to use imputed $X$ for all observations. 

\section{On optimal designs of design-based estimators} \label{optd}

\subsection{Neyman allocation} \label{sec_ney}
Neyman allocation \citep{neyman1934two} minimizes the variance of an estimator of a population mean or total given a fixed sample size. Suppose we want to find the optimal allocation for the population total of the outcome variable $Y$, $T_Y = \sum_{i=1}^{K} N_i \Bar{Y}_i$, where $\Bar{Y}_i$ is the mean of $Y$ for stratum $i$. The objective can be written as
$$
\text{minimize} \hspace{0.3cm} \text{var} \left(T_Y\right) = \sum_{i=1}^{K}\frac{(N_i - n_i)N_i\sigma_i^2}{n_i}  \hspace{0.3cm} \text{suject to}  \hspace{0.3cm} n_1+ n_2 + \dots +n_K = n.
$$
\cite{neyman1934two} showed the optimal allocation is
\begin{align}
    n_k = \frac{N_k\sigma_k}{\sum_{i=1}^{K} N_i\sigma_i}\label{eq1},
\end{align}
where $\sigma_k$ is the standard deviation of $Y$ for stratum $k$. The Equation (\ref{eq1}) does not give integer solutions, and the usual practice is to round off to the nearest integer, but rounding does not necessarily end up with the optimal solution. \cite{wright2017exact} worked out an exact integer algorithm for Neyman allocation, which yields minimum sampling variance.

\subsection{Optimal design for analysis via IPW estimator}\label{secipw}
We are interested in minimizing the variance of $\hat{\beta}_1$. According to Equation (\ref{inf}), the objective becomes
$$
\text{minimize} \hspace{0.3cm} \text{var}\left(\sum_{i=1}^{N}\mathbf{h}_i(\beta_1) \right)   \hspace{0.3cm} \text{suject to}  \hspace{0.3cm} n_1+ n_2 + \dots +n_k = n,
$$
and then optimal allocation for analysis via the IPW estimator is to apply influence functions to Equation (\ref{eq1}), which gives
\begin{align}
    n_{k} \propto N_{k} \sqrt{\operatorname{var}\left(\mathbf{h}_i(\beta_1) \mid \text {stratum } k\right)} \label{eq2}.
\end{align}
\cite{mcisaac2015adaptive} and \cite{amorim2020two} derived similar results by minimizing the asymptotic variance of the mean score and IPW estimator respectively. \cite{chen2020multi} approximated the optimal design (Equation ($\ref{eq2}$)) using a multiwave sampling framework, and their final statistical analyses were conducted using generalized raking estimators. However, Equation (\ref{eq1}) and (\ref{eq2}) need not be the optimal allocation for analysis by generalized raking estimators. It is of interest to find the optimal design for analysis via generalized raking estimators as these are what will be used in analysis.

\subsection{Optimal design for analysis via generalized raking estimators} \label{secraking}
\cite{deville1992calibration} showed all generalized raking estimators are asymptotically equivalent to the generalized regression estimator (GREG), so the optimal design for analysis via a particular generalized raking estimator is also optimal for others. We choose the GREG estimator because it makes arguments more straightforward. Suppose auxiliary variables are $S$, and parameters in the regression estimator are $\theta$. We can then decompose $Y$ as 
$$
Y=(Y- S\theta)+S \theta,
$$
the GREG estimator can then be written as  
$$
\hat{T}=\sum_{i=1}^{N} \frac{R_{i}}{\pi_{i}}\left(Y_{i}-S_{i} \theta\right)+\sum_{i=1}^{N} S_{i} \theta,
$$
where the first term is the IPW estimator of the total of the residuals from regressing $Y$ on $S$. The second term involves the whole population. It has zero variance for any fixed $\theta$ and has variance of smaller order than the first term for any estimator $\theta$ that converges at $\sqrt{n}$ rate. Therefore, to first order, we have
\begin{align*}
    \operatorname{var}\left(\hat{T}\right)=\operatorname{var}\left(\sum_{i=1}^N \frac{R_{i}}{\pi_{i}}\left(Y_{i}-S_{i} \theta\right)\right). 
\end{align*}
The variance of generalized raking estimators of the total is then the variance of the IPW estimator of the total of the residuals $Y_i - S_i\theta$ from regressing $Y$ on $S$. According to Section \ref{sec_ney}, the optimal allocation for analysis via generalized raking estimators is to apply Neyman allocation to residuals, which becomes
\begin{align}
n_{k} \propto N_{k} \sqrt{\operatorname{var}\left[\left(Y_i - S_i\theta \right) \mid \text {stratum } k\right]}. \label{eqgn}
\end{align}

Expression (\ref{eqgn}) poses a problem for the influence-function approach discussed in Section \ref{secipw}. For the influence-function approach, $Y_i$ are the influence functions $\mathbf{h}_i(\beta_1)$ and $S_i$ are the best estimates $\mathbf{h}_i(\hat{\beta}_1)$ we have of them. Let $\gamma$ be the regression parameter from regressing $\mathbf{h}_i(\beta_1)$ on $\mathbf{h}_i(\hat{\beta}_1)$ and $r_i$ be the residuals $r_i =  \mathbf{h}_i(\beta_1) - \mathbf{h}_i(\hat{\beta}_1)\gamma$. The objective becomes 
$$
\text{minimize} \hspace{0.3cm} \text{var} \left( \sum_{i=1}^{N} r_i \right)   \hspace{0.3cm} \text{suject to}  \hspace{0.3cm} n_1+ n_2 + \dots +n_k = n.
$$

A plug-in estimator \citep{kulich2004improving} would estimate the residuals $r_i$ as zero. Before sampling any phase-2 data, we cannot estimate influence functions and residuals. In practice, it is still possible to estimate the residuals using a multiwave sampling framework. After wave 1, $\mathbf{h}_i(\beta_1)$ and $\mathbf{h}_i(\hat{\beta}_1)$ can be estimated using data from the current and previous wave respectively. Besides, We can also simulate to see how optimal design varies, where $\mathbf{h}_i(\beta_1)$ and $\mathbf{h}_i(\hat{\beta}_1)$ can be estimated from the full data and parametric models respectively.

\section{Gain from optimising the design} \label{sims}
Our target is to improve the efficiency of design-based estimators by optimizing the design. However, for generalized raking estimators, if auxiliary variables are good, there is not much room for improvement (without model assumption) because the variance of residuals is small. If auxiliary variables are bad, generalized raking will not improve on the IPW estimator. In this section, we examine the extent to which improvement is still possible.

\subsection{Analytical results}

In this subsection, we compare the optimal design for analysis by the IPW estimator with those for analysis by generalized raking estimators in the classical measurement-error setting where a classical measurement-error model is available and assumed to be correct. Let $\Tilde{X} = X+U$ be the surrogate variables and unbiased measures of $X$, where $U$ has the mean of zero and constant variance of $\sigma^2_U$. 
If $X$ have the mean of $0$, at $\beta = 0$, conditional variance of residuals $r_i$ given stratum $k$ can be written as

\begin{equation}\label{strafo1}
    \mathrm{var}\left(r_i|\text {stratum } k\right)  = \frac{\mathrm{var} \left(U\left(Y-\mu\right)| \text {stratum } k \right)}{\mathrm{var} \left(X\left(Y - \mu \right) |\text {stratum } k \right)}{\mathrm{var}\left(\mathbf{h}_i(0)|\text {stratum } k \right)}.
\end{equation}
The proof is provided in Appendix \ref{proof1}. If we further stratify on $Y$, Equation (\ref{strafo1}) can be simplified to 
\begin{equation}\label{strafo2}
    \mathrm{var}\left(r_i|\text {stratum } k\right)  = \frac{\mathrm{var} \left(U  \right)}{\mathrm{var} \left(X \right)}{\mathrm{var}\left(\mathbf{h}_i(0)|\text {stratum }k \right)}. 
\end{equation}

As $\text{var}(U)/\text{var}(X)$ is a constant across strata, the optimal design will be the same for the two classes of design-based estimators. An important example of outcome-dependent sampling is the case-control study. If we have a rare disease and small covariate effects, the optimal case-control design for analysis by the IPW estimator will sample the same number of cases and controls at $\beta = 0$, whose proof is provided in Appendix \ref{case-control}. According to Equation (\ref{strafo2}), in the same setting, the optimal case-control design for analysis by generalized raking estimators will also take the same number of cases and controls. 

We develop an interactive \texttt{Shiny} app to compare the optimal allocations for analysis by the two classes of design-based estimators, which is available from \url{https://tchen.shinyapps.io/raking}. If we stratify on $Y$ and have small covariate effects, finite-sample numerical results obtained from the \texttt{Shiny} app are consistent with the analytical results of Equation (\ref{strafo2}).

\subsection{Simulation studies} 

In this subsection, we performed extensive simulation studies to compare the efficiencies of different two-phase sampling designs. We also examined the stratum-specific optimal sampling allocations for analysis by the two classes of design-based estimators.

In the first series of simulation studies, we considered an example that the optimal design for analysis by generalized raking estimators was expected to be different from those for analysis by the IPW estimator. We assumed $X$ followed a standard normal distribution, and $Z$ was a binary variable generated from $\text{Bern}(0.5)$. Let $Y$ be the outcome which was generated from the linear model $Y = \beta_0 + \beta_1 X + \beta_2 Z + \epsilon$, where $\epsilon$ followed a standard normal distribution. An auxiliary variable $\mathbf{h}_i^{*}(\beta_1)$, which correlated with the true influence functions $\mathbf{h}_i(\beta_1)$, was simulated with correlation $\rho = \mathrm{cor}(\mathbf{h}_i^{*}(\beta_1), \mathbf{h}_i(\beta_1))$. We set phase-1 sample size $N = 4000$ and phase-2 sample size $n = 600$. The data were stratified into 2 strata based on $\mathbf{h}_i^{*}(\beta_1)$. Specifically, individual $i$ was in stratum $1$ if $\mathbf{h}_i^{*}(\beta_1)$ was in between its $35$th and $65$th percentile and was in stratum $2$ otherwise. We implemented a simple random sampling (SRS), a balanced stratified sampling (BSS) (i.e., $n_1 = n_2 = \dots = n_k$), a proportional stratified sampling (PSS) (i.e., $n_i \propto N_i$), an optimal design for analysis by the IPW estimator where the true influence functions $\mathbf{h}_i(\beta_1)$ were calculated using the (in practice, unavailable) full data (IF-IPW), and an optimal design for analysis by generalized raking estimators where $\mathbf{h}_i(\beta_1)$ and $\mathbf{h}_i^{*}(\beta_1)$ were used as influence functions and their best estimates respectively (IF-GR). We performed the IPW and generalized raking estimations. We used $\mathbf{h}_i^{*}(\beta_1)$ as auxiliary variables in generalized raking procedures. We set $\beta_0 = \beta_1 = \beta_2 = 1$ and performed 2000 Monte Carlo simulations.

For a linear model, influence functions can be written as 
\begin{equation}\label{rak-linear}
    \mathbf{h}_i(\beta) = I^{-1} X^T(Y - \hat{Y}) \approx  I^{-1} X^T\epsilon,
\end{equation}
where $\hat{Y} = X\hat{\beta}$ and $I$ was the per observation population information. According to Equation (\ref{rak-linear}), for the same $X$ and $\epsilon$, varying $\beta$ would not change the true influence functions $\mathbf{h}_i(\beta_1)$. If the stratification was independent of $\beta$, then varying $\beta$ would not change the IF-IPW design. In the first series of simulation studies, stratification and auxiliary variables were independent of regression coefficients $\beta$, so that varying $\beta$ neither changed the IF-IPW design nor the $\hat{\beta}$ estimated from generalized raking.

\begin{table}[H]
\caption{Mean squared error (MSE) and empirical standard error of $\hat{\beta}_1$ estimated from the IPW and generalized raking estimators based on $2000$ Monte Carlo simulations.}
\centering
\begin{tabular}[t]{ccccccccccc}
\toprule
\multicolumn{1}{c}{ } & \multicolumn{2}{c}{SRS} & \multicolumn{2}{c}{BSS} & \multicolumn{2}{c}{PSS} & \multicolumn{2}{c}{IF-IPW} & \multicolumn{2}{c}{IF-GR} \\
\cmidrule(l{3pt}r{3pt}){2-3} \cmidrule(l{3pt}r{3pt}){4-5} \cmidrule(l{3pt}r{3pt}){6-7} \cmidrule(l{3pt}r{3pt}){8-9} \cmidrule(l{3pt}r{3pt}){10-11}
$\rho$ & MSE* & se & MSE* & se & MSE* & se & MSE* & se & MSE* & se\\
\midrule
\addlinespace[0.3em]
\multicolumn{11}{l}{\textbf{IPW}}\\
\hspace{1em}0.99 & 1.65 & 0.041 & 2.31 & 0.048 & 1.66 & 0.041 & 1.36 & 0.037 & 1.63 & 0.040\\
\hspace{1em}0.90 & 1.67 & 0.041 & 2.27 & 0.048 & 1.64 & 0.041 & 1.43 & 0.038 & 1.54 & 0.039\\
\hspace{1em}0.80 & 1.66 & 0.041 & 2.37 & 0.049 & 1.67 & 0.041 & 1.52 & 0.039 & 1.63 & 0.040\\
\hspace{1em}0.70 & 1.76 & 0.042 & 2.25 & 0.047 & 1.64 & 0.041 & 1.58 & 0.040 & 1.62 & 0.040\\
\hspace{1em}0.60 & 1.60 & 0.040 & 2.29 & 0.048 & 1.72 & 0.042 & 1.65 & 0.041 & 1.55 & 0.039\\
\hspace{1em}0.50 & 1.67 & 0.041 & 2.04 & 0.045 & 1.66 & 0.041 & 1.71 & 0.041 & 1.67 & 0.041\\
\addlinespace[0.3em]
\multicolumn{11}{l}{\textbf{Generalized Raking}}\\
\hspace{1em}0.99 & 0.29 & 0.017 & 0.30 & 0.017 & 0.29 & 0.017 & 0.33 & 0.018 & 0.28 & 0.017\\
\hspace{1em}0.90 & 0.55 & 0.024 & 0.62 & 0.025 & 0.53 & 0.023 & 0.57 & 0.024 & 0.50 & 0.022\\
\hspace{1em}0.80 & 0.79 & 0.028 & 0.98 & 0.031 & 0.78 & 0.028 & 0.84 & 0.029 & 0.81 & 0.028\\
\hspace{1em}0.70 & 1.04 & 0.032 & 1.27 & 0.036 & 1.00 & 0.032 & 1.03 & 0.032 & 0.96 & 0.031\\
\hspace{1em}0.60 & 1.15 & 0.034 & 1.50 & 0.039 & 1.19 & 0.035 & 1.17 & 0.034 & 1.16 & 0.034\\
\hspace{1em}0.50 & 1.35 & 0.037 & 1.57 & 0.040 & 1.34 & 0.037 & 1.35 & 0.037 & 1.33 & 0.036\\
\bottomrule
\end{tabular}
\begin{tablenotes}
\item Note: MSE*: MSE$\times 1000$; se: empirical standard error of $\hat{\beta}_1$.
\end{tablenotes}
\label{rak_table_1}
\end{table}

Results of the first series of simulation studies were given in Table \ref{rak_table_1} and Supplementary Figure 1. For analysis results of the IPW estimator, when $\mathbf{h}_i^{*}(\beta_1)$ were highly correlated with $\mathbf{h}_i(\beta_1)$, IF-IPW was more efficient than other sampling designs. The efficiency gain decreased as the correlation $\rho$ decreased. The gain was not obvious when $\mathbf{h}_i^{*}(\beta_1)$ were not good approximations of $\mathbf{h}_i(\beta_1)$.  BSS was the least efficient design as it sampled a lot more people from stratum $1$ compared with the IF-IPW design. PPS was as efficient as SRS, and IF-GR was slightly more efficient than them. The results seconded that IF-GR was not the optimal design for analysis by the IPW estimator.

Analysis results of generalized raking estimators showed that generalized raking estimators were much more efficient than the IPW estimator. The efficiency decreased as $\rho$ decreased for all five designs. As expected, when $\rho = 0.99$, all five designs ended up with similar efficiencies because there was not much room for gaining from optimizing the design. When $\rho \neq 0.99$, IF-GR was slightly more efficient than IF-IPW, PSS, and SRS. BSS was the least efficient design. The results showed that IF-IPW was not optimal for analysis by generalized raking estimators, though it ended up with a very similar efficiency as IF-GR. Supplementary Figure 1 showed that IF-GR could be very different from IF-IPW. IF-GR sampled more individuals from the stratum 1 (the middle of the distribution of $\mathbf{h}_i^{*}(\beta_1)$) compared with IF-IPW. As $\rho$ increased, the two designs would be more different. 

Table \ref{rak_table_1} showed IF-GR was also as efficient as SRS and PSS, which was partly due to the variances of residuals $\mathrm{var}(r_i)$ were roughly constant across strata, so that these three designs were close. If $\mathrm{var}(r_i)$ were more different across strata, we expected IF-GR would be different and more efficient than SRS and PSS. Based on data simulated from the first series of simulation studies, we further simulated an auxiliary variable $\mathbf{h}_i^{\star}(\beta_1)$ which would be used in the IF-GR design and generalized raking analysis.  We set the tail stratum variance larger and kept the correlation between $\mathbf{h}_i(\beta_1)$ and $\mathbf{h}_i^{\star}(\beta_1)$ around $0.7$. Specifically, let $\mathbf{h}_i^{\star}(\beta_1) = \mathbf{h}_i(\beta_1) - 3\times r_i$ if individual $i$ was in stratum $2$ and $\mathbf{h}_i^{\star}(\beta_1) = \mathbf{h}_i(\beta_1) - r_i$ otherwise. Results were given in Table \ref{rak_table_2}. As we made the variance of residuals of stratum $2$ larger, IF-GR would sample more individuals from stratum $2$, so IF-GR and IF-IPW would be closer. As expected, IF-IPW was as efficient as IF-GR, and they were considerably more efficient than the other three designs.

\begin{table}
\caption{Mean squared error (MSE) and empirical standard error of $\hat{\beta}_1$ estimated from generalized raking estimators based on $2000$ Monte Carlo simulations.}
\centering
\begin{tabular}[t]{rrrrrrrrrr}
\toprule
\multicolumn{2}{c}{SRS} & \multicolumn{2}{c}{BSS} & \multicolumn{2}{c}{PSS} & \multicolumn{2}{c}{IF-IPW} & \multicolumn{2}{c}{IF-GR} \\
\cmidrule(l{3pt}r{3pt}){1-2} \cmidrule(l{3pt}r{3pt}){3-4} \cmidrule(l{3pt}r{3pt}){5-6} \cmidrule(l{3pt}r{3pt}){7-8} \cmidrule(l{3pt}r{3pt}){9-10}
MSE* & se & MSE* & se & MSE* & se & MSE* & se & MSE* & se\\
\midrule
0.95 & 0.031 & 1.29 & 0.036 & 0.96 & 0.031 & 0.85 & 0.029 & 0.83 & 0.029\\
\bottomrule
\end{tabular}
\begin{tablenotes}
\item Note: MSE*: MSE$\times 1000$; se: empirical standard error of $\hat{\beta}_1$.
\end{tablenotes}
\label{rak_table_2}
\end{table}

We then conducted simulation studies in measurement-error settings. In the second series of simulation studies, we considered the situation that both variables of interest $X$ and outcome $Y$ were continuous. We assumed $X$ followed a standard normal distribution. The error-prone variable $\Tilde{X}$ was assumed to have additive error $\Tilde{X} = X + U$, where $U \sim N(0, \sigma^2)$. The outcome of interest $Y$ was generated from the linear model $Y = \beta_0 + \beta_1 X + \beta_2 Z_1 + \beta_3 Z_2 + \epsilon$, where $Z_1 \sim \mathrm{Bern}(0.5)$, $Z_2$ followed a standard uniform distribution, and $\epsilon$ followed a standard normal distribution. The data were divided into 3 strata based on $\Tilde{X}$ ($\leq20$th, $>20$th to $\leq80$th, and $>80$th percentiles).  We set $N = 4000$, $n = 600$, and $\beta_0 = \beta_2 = \beta_3 = \beta_4 = 1$. We considered the same sampling strategies as those implemented in the first series of simulation studies. The only difference was that, for the IF-GR design, the best estimates of influence functions were calculated using multiple imputation with 50 imputations. Specifically, imputations were calculated by Bayesian linear regression using the \texttt{mice} package \citep{mice2011} in \texttt{R}.  We performed the IPW and generalized raking estimations. The procedures described in Section \ref{gr-procedure} were adopted to calculate auxiliary variables in the final analysis.

\begin{table}[H]
\caption{Mean squared error (MSE) and empirical standard error of $\hat{\beta}_1$ estimated from the IPW and generalized raking estimators for linear regression with continuous $X$ based on $2000$ Monte Carlo simulations.}
\centering
\begin{tabular}[t]{cccccccccccc}
\toprule
\multicolumn{1}{c}{ } & \multicolumn{1}{c}{ } & \multicolumn{2}{c}{SRS} & \multicolumn{2}{c}{BSS} & \multicolumn{2}{c}{PSS} & \multicolumn{2}{c}{IF-IPW} & \multicolumn{2}{c}{IF-GR} \\
\cmidrule(l{3pt}r{3pt}){3-4} \cmidrule(l{3pt}r{3pt}){5-6} \cmidrule(l{3pt}r{3pt}){7-8} \cmidrule(l{3pt}r{3pt}){9-10} \cmidrule(l{3pt}r{3pt}){11-12}
$\sigma$ & $\beta_1$ & MSE* & se & MSE* & se & MSE* & se & MSE* & se & MSE* & se\\
\midrule
\addlinespace[0.3em]
\multicolumn{12}{l}{\textbf{IPW}}\\
\hspace{1em} & 0 & 1.58 & 0.040 & 1.51 & 0.039 & 1.67 & 0.041 & 1.42 & 0.038 & 1.54 & 0.039\\

\hspace{1em} & 1 & 1.73 & 0.042 & 1.42 & 0.038 & 1.65 & 0.041 & 1.42 & 0.038 & 1.49 & 0.039\\

\hspace{1em}\multirow[t]{-3}{*}{\raggedleft\arraybackslash 0.50} & 2 & 1.68 & 0.041 & 1.42 & 0.038 & 1.63 & 0.040 & 1.43 & 0.038 & 1.46 & 0.038\\

\hspace{1em} & 0 & 1.66 & 0.041 & 1.50 & 0.039 & 1.62 & 0.040 & 1.58 & 0.040 & 1.64 & 0.040\\

\hspace{1em} & 1 & 1.71 & 0.041 & 1.73 & 0.042 & 1.71 & 0.041 & 1.54 & 0.039 & 1.57 & 0.040\\

\hspace{1em}\multirow[t]{-3}{*}{\raggedleft\arraybackslash 0.75} & 2 & 1.62 & 0.040 & 1.77 & 0.042 & 1.70 & 0.041 & 1.54 & 0.039 & 1.60 & 0.040\\

\hspace{1em} & 0 & 1.69 & 0.041 & 1.65 & 0.041 & 1.66 & 0.041 & 1.54 & 0.039 & 1.79 & 0.042\\

\hspace{1em} & 1 & 1.76 & 0.042 & 1.74 & 0.042 & 1.67 & 0.041 & 1.56 & 0.039 & 1.57 & 0.040\\

\hspace{1em}\multirow[t]{-3}{*}{\raggedleft\arraybackslash 1.00} & 2 & 1.77 & 0.042 & 1.71 & 0.041 & 1.74 & 0.042 & 1.57 & 0.040 & 1.60 & 0.040\\

\addlinespace[0.3em]
\multicolumn{12}{l}{\textbf{Generalized Raking}}\\
\hspace{1em} & 0 & 0.53 & 0.023 & 0.68 & 0.026 & 0.53 & 0.023 & 0.58 & 0.024 & 0.54 & 0.023\\

\hspace{1em} & 1 & 0.75 & 0.027 & 0.78 & 0.028 & 0.71 & 0.027 & 0.72 & 0.027 & 0.74 & 0.027\\

\hspace{1em}\multirow[t]{-3}{*}{\raggedleft\arraybackslash 0.50} & 2 & 1.03 & 0.032 & 1.00 & 0.032 & 0.98 & 0.031 & 0.99 & 0.031 & 0.98 & 0.031\\

\hspace{1em} & 0 & 0.79 & 0.028 & 0.91 & 0.030 & 0.79 & 0.028 & 0.81 & 0.028 & 0.78 & 0.028\\

\hspace{1em} & 1 & 1.07 & 0.033 & 1.17 & 0.034 & 1.00 & 0.032 & 1.02 & 0.032 & 0.99 & 0.032\\

\hspace{1em}\multirow[t]{-3}{*}{\raggedleft\arraybackslash 0.75} & 2 & 1.27 & 0.036 & 1.41 & 0.038 & 1.32 & 0.036 & 1.30 & 0.036 & 1.31 & 0.036\\

\hspace{1em} & 0 & 0.96 & 0.031 & 1.15 & 0.034 & 0.95 & 0.031 & 1.02 & 0.032 & 0.97 & 0.031\\

\hspace{1em} & 1 & 1.26 & 0.036 & 1.43 & 0.038 & 1.18 & 0.034 & 1.16 & 0.034 & 1.20 & 0.035\\

\hspace{1em}\multirow[t]{-3}{*}{\raggedleft\arraybackslash 1.00} & 2 & 1.47 & 0.038 & 1.57 & 0.040 & 1.52 & 0.039 & 1.38 & 0.037 & 1.44 & 0.038\\
\bottomrule
\end{tabular}
\begin{tablenotes}
\item Note: MSE*: MSE$\times 1000$; se: empirical standard error of $\hat{\beta}_1$.
\end{tablenotes}
\label{rak_table_3}
\end{table}

Results of the second series of simulation studies were given in Table \ref{rak_table_3} and Supplementary Figure 2 -- 4. The results were consistent with the previous simulation studies. For the analysis results of the IPW estimator, IF-IPW was more efficient than other designs. The efficiency gain decreased as $\sigma$ increased. For the analysis results of generalized raking estimators, when $\sigma = 0.5$, all five designs are equally efficient as $\Tilde{X}$ was a good surrogate of $X$. When $\sigma$ got larger, IF-GR was as efficient as IF-IPW, SRS, and PSS. BSS was the least efficient design compared with the other designs. For all five designs, the efficiency decreased as $\sigma$ and $\beta$ increased. Supplementary Figure 2 -- 4 showed IF-GR tended to sample more observations from the middle stratum compared with IF-IPW, though it would be closer to IF-IPW as $\beta$ and $\sigma$ increased.

In the third series of simulation studies, we let the outcome $Y$ be binary and variables of interest $X$ be continuous. $X$, $Z_1$, and $\Tilde{X}$ were generated exactly the same as the second series of simulation studies. The binary outcome was generated from $P(Y|X,Z_1) = \mathrm{expit}(\beta_0 + \beta_1 X + \beta_2 Z_1)$, where $\mathrm{expit}(x) = \mathrm{exp}(x)/(1+\mathrm{exp}(x))$. The data were stratified into 6 strata base on $Y$ and the first and third quartile of $\Tilde{X}$. We set $\beta_0 = -1.5$, $\beta_2 = 1$, $N =4000$, and $n = 600$. We considered the same sampling strategies as those implemented in the previous simulation studies.

Results of the third series of simulation studies were shown in Table \ref{rak_table_4} and Supplementary Figure 5 -- 7. For analysis results of the IPW estimator, IF-IPW was as efficient as IF-GR. When $\beta_1 = 0$, BSS was also as efficient as IF-IPW and IF-GR. The performance of BSS got worse as $\beta_1$ increased. Typically, PSS and SRS were less efficient compared with the other three designs. For analysis results of generalized raking estimators, we observed IF-IPW and IF-GR had similar performance and were more efficient than other designs. Supplementary Figure 5 -- 7 showed that compared with a continuous outcome, IF-GR was closer to IF-IPW for a binary outcome, though IF-GR also sampled more people from the middle of the distribution of $\Tilde{X}$ in each stratum defined by $Y$.

\begin{table}[H]
\caption{Mean squared error (MSE) and empirical standard error of $\hat{\beta}_1$ estimated from the IPW and generalized raking estimators for logistic regression with continuous $X$ based on $2000$ Monte Carlo simulations.}
\centering
\begin{tabular}[t]{cccccccccccc}
\toprule
\multicolumn{1}{c}{ } & \multicolumn{1}{c}{ } & \multicolumn{2}{c}{SRS} & \multicolumn{2}{c}{BSS} & \multicolumn{2}{c}{PSS} & \multicolumn{2}{c}{IF-IPW} & \multicolumn{2}{c}{IF-GR} \\
\cmidrule(l{3pt}r{3pt}){3-4} \cmidrule(l{3pt}r{3pt}){5-6} \cmidrule(l{3pt}r{3pt}){7-8} \cmidrule(l{3pt}r{3pt}){9-10} \cmidrule(l{3pt}r{3pt}){11-12}
$\sigma$ & $\beta_1$ & MSE* & se & MSE* & se & MSE* & se & MSE* & se & MSE* & se\\
\midrule
\addlinespace[0.3em]
\multicolumn{12}{l}{\textbf{IPW}}\\
\hspace{1em} & 0.0 & 3.86 & 0.062 & 3.28 & 0.057 & 3.69 & 0.061 & 3.35 & 0.058 & 3.26 & 0.057\\

\hspace{1em} & 0.5 & 4.47 & 0.067 & 3.96 & 0.063 & 4.65 & 0.068 & 3.61 & 0.060 & 3.64 & 0.060\\

\hspace{1em}\multirow[t]{-3}{*}{\raggedleft\arraybackslash 0.50} & 1.0 & 7.76 & 0.088 & 6.16 & 0.078 & 7.03 & 0.083 & 4.95 & 0.070 & 5.25 & 0.072\\

\hspace{1em} & 0.0 & 4.80 & 0.069 & 4.42 & 0.067 & 4.79 & 0.069 & 4.02 & 0.063 & 4.14 & 0.064\\

\hspace{1em} & 0.5 & 5.80 & 0.076 & 5.29 & 0.073 & 5.38 & 0.073 & 4.60 & 0.068 & 4.49 & 0.067\\

\hspace{1em}\multirow[t]{-3}{*}{\raggedleft\arraybackslash 0.75} & 1.0 & 9.25 & 0.096 & 7.79 & 0.088 & 8.84 & 0.094 & 6.50 & 0.081 & 5.97 & 0.077\\

\hspace{1em} & 0.0 & 5.81 & 0.076 & 5.10 & 0.071 & 5.83 & 0.076 & 4.92 & 0.070 & 4.71 & 0.069\\

\hspace{1em} & 0.5 & 6.68 & 0.082 & 6.38 & 0.080 & 6.93 & 0.083 & 5.63 & 0.075 & 5.64 & 0.075\\

\hspace{1em}\multirow[t]{-3}{*}{\raggedleft\arraybackslash 1.00} & 1.0 & 10.16 & 0.101 & 9.59 & 0.098 & 10.13 & 0.101 & 7.41 & 0.086 & 8.17 & 0.090\\

\addlinespace[0.3em]
\multicolumn{12}{l}{\textbf{Generalized Raking}}\\
\hspace{1em} & 0.0 & 2.33 & 0.048 & 2.13 & 0.046 & 2.29 & 0.048 & 2.05 & 0.045 & 2.08 & 0.046\\

\hspace{1em} & 0.5 & 2.84 & 0.053 & 2.52 & 0.050 & 2.74 & 0.052 & 2.33 & 0.048 & 2.33 & 0.048\\

\hspace{1em}\multirow[t]{-3}{*}{\raggedleft\arraybackslash 0.50} & 1.0 & 5.06 & 0.071 & 4.11 & 0.064 & 4.71 & 0.068 & 3.48 & 0.059 & 3.46 & 0.059\\

\hspace{1em} & 0.0 & 3.53 & 0.059 & 3.27 & 0.057 & 3.47 & 0.059 & 3.01 & 0.055 & 3.05 & 0.055\\

\hspace{1em} & 0.5 & 4.45 & 0.067 & 4.08 & 0.064 & 4.08 & 0.064 & 3.61 & 0.060 & 3.50 & 0.059\\

\hspace{1em}\multirow[t]{-3}{*}{\raggedleft\arraybackslash 0.75} & 1.0 & 7.32 & 0.086 & 6.44 & 0.080 & 7.07 & 0.084 & 5.38 & 0.073 & 5.03 & 0.071\\

\hspace{1em} & 0.0 & 4.83 & 0.070 & 4.41 & 0.066 & 4.80 & 0.069 & 3.89 & 0.062 & 3.89 & 0.062\\

\hspace{1em} & 0.5 & 5.61 & 0.075 & 5.49 & 0.074 & 5.77 & 0.076 & 4.91 & 0.070 & 4.81 & 0.069\\

\hspace{1em}\multirow[t]{-3}{*}{\raggedleft\arraybackslash 1.00} & 1.0 & 8.79 & 0.094 & 8.80 & 0.094 & 8.72 & 0.093 & 6.42 & 0.080 & 7.04 & 0.084\\
\bottomrule
\end{tabular}
\begin{tablenotes}
\item Note: MSE*: MSE$\times 1000$; se: empirical standard error of $\hat{\beta}_1$.
\end{tablenotes}
\label{rak_table_4}
\end{table}

In the last series of simulation studies, we considered that the outcome $Y$, variables of interest $X$, and covariates $Z_1$ were all binary. Let $X \sim \mathrm{Bern}(0.4)$ and $Z_1 \sim \mathrm{Bern}(0.5)$. An error-prone variable $\Tilde{X}$ was generated with prespecified sensitivity and specificity. The binary outcome was generated from $P(Y|X,Z_1) = \mathrm{expit}(\beta_0 - \beta_1 X + \beta_2 Z_1)$. We considered a rare disease with $E(Y) = 0.05$ which was controlled by $\beta_0$. We set $\beta_2 = 1$ and $N=10000$. The data were divided into 4 strata based on $Y$ and $\Tilde{X}$. Instead of balanced stratified sampling, we implemented a stratified case-control sampling (SCC) which sampled all cases and the same number of controls in each stratum defined by $\Tilde{X}$.

Results of the last series of simulation studies were shown in Table \ref{rak_table_5} and Supplementary Figure 8 -- 10. For analysis results of the IPW estimator, IF-IPW was as efficient as IF-GR. They were slightly more efficient than SCC and much more efficient than PSS and SRS. For analysis results of generalized raking estimators, IF-IPW and IF-GR are equally efficient. SCC was also close to them. These three designs were substantially more efficient than PSS and SRS. Supplementary Figure 8 -- 10 showed IF-GR was quite close to IF-IPW when data were all discrete. This was consistent with the findings of \cite{mcisaac2014response}. 

\begin{table}[H]
\caption{Mean squared error (MSE) and empirical standard error of $\hat{\beta}_1$ estimated from the IPW and generalized raking estimators for logistic regression with binary $X$ based on $2000$ Monte Carlo simulations.}
\centering
\begin{tabular}[t]{ccccccccccccc}
\toprule
\multicolumn{1}{c}{ } & \multicolumn{1}{c}{ } & \multicolumn{1}{c}{ } & \multicolumn{2}{c}{SRS} & \multicolumn{2}{c}{SCC} & \multicolumn{2}{c}{PSS} & \multicolumn{2}{c}{IF-IPW} & \multicolumn{2}{c}{IF-GR} \\
\cmidrule(l{3pt}r{3pt}){4-5} \cmidrule(l{3pt}r{3pt}){6-7} \cmidrule(l{3pt}r{3pt}){8-9} \cmidrule(l{3pt}r{3pt}){10-11} \cmidrule(l{3pt}r{3pt}){12-13}
Sen & Spe & $\beta_1$ & MSE$^{\star}$ & se & MSE$^{\star}$ & se & MSE$^{\star}$ & se & MSE$^{\star}$ & se & MSE$^{\star}$ & se\\
\midrule
\addlinespace[0.3em]
\multicolumn{13}{l}{\textbf{IPW}}\\
\hspace{1em} &  & 0.0 & 2.70 & 0.164 & 1.21 & 0.110 & 2.52 & 0.159 & 1.18 & 0.109 & 1.18 & 0.109\\

\hspace{1em} &  & 0.5 & 3.33 & 0.182 & 1.32 & 0.115 & 3.09 & 0.176 & 1.21 & 0.110 & 1.24 & 0.111\\

\hspace{1em}\multirow[t]{-3}{*}{\raggedleft\arraybackslash 0.95} & \multirow[t]{-3}{*}{\raggedleft\arraybackslash 0.95} & 1.0 & 5.01 & 0.223 & 1.67 & 0.129 & 4.80 & 0.219 & 1.53 & 0.124 & 1.54 & 0.124\\

\hspace{1em} &  & 0.0 & 4.14 & 0.204 & 1.35 & 0.116 & 4.06 & 0.202 & 1.37 & 0.117 & 1.35 & 0.116\\

\hspace{1em} &  & 0.5 & 5.63 & 0.237 & 1.49 & 0.122 & 5.02 & 0.224 & 1.44 & 0.120 & 1.47 & 0.121\\

\hspace{1em}\multirow[t]{-3}{*}{\raggedleft\arraybackslash 0.90} & \multirow[t]{-3}{*}{\raggedleft\arraybackslash 0.90} & 1.0 & 7.92 & 0.280 & 1.95 & 0.139 & 7.31 & 0.270 & 1.67 & 0.129 & 1.76 & 0.133\\

\hspace{1em} &  & 0.0 & 5.34 & 0.231 & 1.49 & 0.122 & 5.07 & 0.225 & 1.44 & 0.120 & 1.42 & 0.119\\

\hspace{1em} &  & 0.5 & 6.91 & 0.263 & 1.66 & 0.129 & 6.57 & 0.256 & 1.55 & 0.124 & 1.53 & 0.124\\

\hspace{1em}\multirow[t]{-3}{*}{\raggedleft\arraybackslash 0.85} & \multirow[t]{-3}{*}{\raggedleft\arraybackslash 0.85} & 1.0 & 9.26 & 0.304 & 2.07 & 0.144 & 9.70 & 0.311 & 1.83 & 0.135 & 1.79 & 0.134\\

\addlinespace[0.3em]
\multicolumn{13}{l}{\textbf{Generalized Raking}}\\
\hspace{1em} &  & 0.0 & 2.65 & 0.163 & 1.06 & 0.103 & 2.51 & 0.158 & 1.07 & 0.104 & 1.05 & 0.103\\

\hspace{1em} &  & 0.5 & 3.29 & 0.181 & 1.16 & 0.108 & 3.03 & 0.174 & 1.14 & 0.107 & 1.14 & 0.107\\

\hspace{1em}\multirow[t]{-3}{*}{\raggedleft\arraybackslash 0.95} & \multirow[t]{-3}{*}{\raggedleft\arraybackslash 0.95} & 1.0 & 4.96 & 0.222 & 1.51 & 0.123 & 4.83 & 0.219 & 1.43 & 0.119 & 1.42 & 0.119\\

\hspace{1em} &  & 0.0 & 4.14 & 0.204 & 1.24 & 0.111 & 4.04 & 0.201 & 1.29 & 0.114 & 1.24 & 0.112\\

\hspace{1em} &  & 0.5 & 5.62 & 0.236 & 1.39 & 0.118 & 4.97 & 0.223 & 1.33 & 0.115 & 1.35 & 0.116\\

\hspace{1em}\multirow[t]{-3}{*}{\raggedleft\arraybackslash 0.90} & \multirow[t]{-3}{*}{\raggedleft\arraybackslash 0.90} & 1.0 & 7.94 & 0.281 & 1.81 & 0.135 & 7.30 & 0.270 & 1.60 & 0.126 & 1.66 & 0.129\\

\hspace{1em} &  & 0.0 & 5.37 & 0.232 & 1.41 & 0.119 & 5.17 & 0.227 & 1.36 & 0.117 & 1.35 & 0.116\\

\hspace{1em} &  & 0.5 & 7.00 & 0.264 & 1.58 & 0.126 & 6.68 & 0.258 & 1.46 & 0.121 & 1.48 & 0.122\\

\hspace{1em}\multirow[t]{-3}{*}{\raggedleft\arraybackslash 0.85} & \multirow[t]{-3}{*}{\raggedleft\arraybackslash 0.85} & 1.0 & 9.37 & 0.305 & 1.96 & 0.140 & 9.94 & 0.314 & 1.77 & 0.133 & 1.75 & 0.132\\
\bottomrule
\end{tabular}
\begin{tablenotes}
\item Note: MSE$^{\star}$: MSE$\times 100$; se: empirical standard error of $\hat{\beta}_1$; Sen: sensitivity; Spe: specificity.
\end{tablenotes}
\label{rak_table_5}
\end{table}

\section{Data example: National Wilms' Tumor Study}\label{dataeg}
In this section, we compared different two-phase sampling strategies using the data example from the National Wilms' Tumor Study \citep{d1989treatment, green1998comparison}. The cohort consisted of 3915 patients. Variables available for the whole cohort included histology evaluated by central lab (favorable vs unfavorable (histol)), histology evaluated by institution (favorable vs unfavorable (instit)), age at diagnosis (age), stage of disease (stage), study, diameter of tumor (tumdiam), and an indicator of relapse (relapse). We assumed histology evaluated by central lab (histol) was the phase-2 variable and correlated with histology evaluated by institution. 

The data were stratified into 4 strata based on the indicator of relapse and histology evaluated by institution. The phase-1 sample sizes were  $(3026,220, 507,162)$. Following \cite{kulich2004improving, breslow2009improved, chen2020multi}, we fitted a logistic model with model terms histology (histol), a linear spline with the separate slope for greater or less than 1 year old, a binary indicator of stage (I-II vs. III-IV), and interactions between stage and diameter. We performed the IPW and generalized raking estimations. We followed the procedures described in Section \ref{gr-procedure} and imputed central lab histology using a logistic model with model terms histology evaluated by institution, a binary indicator of age ($>10$ vs. $\leq10$) and interaction between study and a binary indicator of stage (1-III vs. IV).

We considered a stratified case-control sampling (SCC) which sampled all cases and the same number of controls in each histology stratum, a proportional stratified sampling (PSS), an optimal sampling for analysis by the IPW estimator (IF-IPW) where influence functions were computed using the full data, and an optimal sampling for analysis by generalized raking estimators (IF-GR) where the best estimates of the influence functions were calculated by multiple imputation with 50 imputations.

Detailed sampling results were shown in Table \ref{rak_table_wilm_design}. SRS and PSS sampled too few relapsed cases. Compared with SRS and PSS, SCC sampled more relapsed cases and rare unfavourable histology controls. IF-IPW was close to IF-GR, and they sampled a few more favourable histology controls compared with SCC.

Analysis results were shown in Table \ref{rak_table_6}. Generalized raking estimators were more efficient than the IPW estimator. As expected, IF-GR was as efficient as IF-IPW, and they were much more efficient than the other three designs. SCC was also more efficient than SRS and PSS.

\begin{table}[H]
\caption{A comparison of different sampling strategies where the strata were defined based on institutional histology and outcome for Wilm's Tumor.}
\centering
\begin{tabular}[t]{ccccccc}
\toprule
 & Stratum Size & $\overline{\text{SRS}}$ & SCC & PSS & IF-IPW & $\overline{\text{IF-GR}}$\\
\midrule
Controls \& Favourable  & 3026 & 1034 & 507 & 1034 & 736 & 747\\
Controls \& Unfavourable & 220 & 75 & 162 & 75 & 144 & 142\\
Cases \& Favourable & 507 & 173 & 507 & 173 & 345 & 351\\
Cases \& Unfavourable & 162 & 55 & 162 & 55 & 113 & 98\\
\bottomrule
\end{tabular}
\begin{tablenotes}
\item Note: $\overline{\text{SRS}}$: the mean stratum-specific sample size of simple random sampling based on $2000$ Monte Carlo simulations. $\overline{\text{IF-GR}}$: the mean stratum-specific sample size of the optimal sampling for analysis by generalized raking estimators based on $2000$ Monte Carlo simulations.
\end{tablenotes}
\label{rak_table_wilm_design}
\end{table}

\begin{table}[H]
\caption{Mean squared error (MSE) and empirical standard error of $\hat{\beta}_{histol}$ estimated from the IPW and generalized raking estimator based on $2000$ Monte Carlo simulations.}
\centering
\begin{tabular}[t]{cccccccccc}
\toprule
 \multicolumn{2}{c}{SRS} & \multicolumn{2}{c}{SCC} & \multicolumn{2}{c}{PSS} & \multicolumn{2}{c}{IF-IPW} & \multicolumn{2}{c}{IF-GR} \\
\cmidrule(l{3pt}r{3pt}){1-2} \cmidrule(l{3pt}r{3pt}){3-4} \cmidrule(l{3pt}r{3pt}){5-6} \cmidrule(l{3pt}r{3pt}){7-8} \cmidrule(l{3pt}r{3pt}){9-10}
MSE$^{\star}$ & se & MSE$^{\star}$ & se & MSE$^{\star}$ & se & MSE$^{\star}$ & se & MSE$^{\star}$ & se\\
\midrule
\addlinespace[0.3em]
\multicolumn{10}{l}{\textbf{IPW}}\\
\hspace{1em}  1.07 & 0.103 & 0.79 & 0.089 & 1.04 & 0.102 & 0.69 & 0.083 & 0.66 & 0.081\\
\addlinespace[0.3em]
\multicolumn{10}{l}{\textbf{Generalized Raking}}\\
\hspace{1em}  0.97 & 0.098 & 0.77 & 0.088 & 0.93 & 0.097 & 0.65 & 0.081 & 0.63 & 0.079\\
\bottomrule
\end{tabular}
\begin{tablenotes}
\item Note: MSE$^{\star}$: MSE$\times 100$; se: empirical standard error of $\hat{\beta}_{histol}$.
\end{tablenotes}
\label{rak_table_6}
\end{table}

\section{Discussion}\label{final}

In this article, we explored the optimal sampling for design-based estimators of regression models in two-phase designs. We derived a closed-form expression of the optimal design for analysis by generalized raking estimators. For analysis by the IPW estimator, the optimal design is to apply Neyman allocation to influence functions. For analysis by generalized raking estimators, the optimal design is to apply Neyman allocation to residuals from regressing the influences functions $\mathbf{h}_i(\beta_1)$ on their best estimates $\mathbf{h}_i(\hat{\beta}_1)$.

In practice, it is hard to approximate the optimal design for analysis by generalized raking estimators. In order to approximate the design, we need to estimate the influence functions and their best estimates, which cannot be done at the design stage or for a single-wave sampling. However, it is still possible to approximate the design using a multiwave adaptive estimator \citep{mcisaac2015adaptive}. After wave 1, influence functions estimated from the current and previous wave can be used as $\mathbf{h}_i(\beta_1)$ and their best estimates $\mathbf{h}_i(\hat{\beta}_1)$ respectively.

In our simulation studies, proportional stratified sampling and simple random sampling are highly efficient for generalized raking estimation with a continuous outcome. The results are consistent with simulation studies of \cite{amorim2020two}. If we have good auxiliary variables, the residuals from regressing influence functions $\mathbf{h}_i(\beta_1)$ on their best estimates $\mathbf{h}_i(\hat{\beta}_1)$ will be roughly constant, so that the optimal design will be close to proportional stratified sampling and simple random sampling. When data are all discrete, the optimal design for analysis by the IPW estimator is close to those for analysis by generalized raking estimators, which are also discussed by \cite{mcisaac2014response}.

If auxiliary variables are good, $\mathbf{h}_i(\hat{\beta}_1)$ will be highly correlated with $\mathbf{h}_i(\beta_1)$, so there is not much room for gains from optimizing the design. If auxiliary variables are bad, on the one hand, generalized raking estimators will not improve on the IPW estimator; on the other hand, the variance of residuals $\mathrm{var}{(r_i)}$ will be close to the variance of influence functions $\mathrm{var}{(\mathbf{h}_i(\beta_1))}$, so that the optimal design for analysis by the IPW estimator is close to those for analysis by generalized raking estimators. If auxiliary variables are neither too good nor too bad, it may still be possible to gain from optimizing the design.

We showed that optimal designs for analysis by the IPW and generalized raking estimators are different but often have similar efficiency. Lack of improvement is desired in practice because it is hard to approximate the optimal design for generalized raking estimations.

\section*{ACKNOWLEDGMENTS}
This work was supported in part by the University of Auckland doctoral scholarship (to the first author),
 Patient Centered Outcomes Research Institute
(PCORI) Award R-1609-36207 and U.S. National Institutes of Health (NIH) grant R01-
AI131771. The statements in this manuscript are solely the responsibility of the authors and do not necessarily represent the views of PCORI or NIH.

\section*{Appendix}\label{app1}
\appendix
\section{Technical detail} \label{proof1}
The goal is to minimize
$$
 \text{var} \left( \sum_{i=1}^{N} \left( \mathbf{h}_i(\beta) - \mathbf{h}_i(\hat{\beta})\gamma \right) \right)   \hspace{0.3cm} \text{suject to}  \hspace{0.3cm} n_1+ n_2 + \dots +n_k = n.
$$
Suppose $\Tilde{X} = X + U$ are surrogate variables of $X$, where $U$ has mean zero and constant variance $\sigma_U^2$. Regression calibration \citep{prentice1982covariate} suggested we can regress $Y$ on $E(X|\Tilde{X})$ to get unbiased estimates of $\beta$. If $X$ have mean 0, we have
$$
E(X|\Tilde{X}) = \lambda \Tilde{X},
$$
where $\lambda = \text{var} (X)/ \text{var} (\Tilde{X})$. At $\beta = 0$, $\mathbf{h}_i(\beta)$ and $\mathbf{h}_i(\hat{\beta})$ can be estimated by $X(Y-\mu)$ and $\lambda \Tilde{X} (Y-\mu)$ respectively, where $\mu$ is a constant. Since $X$ are independent of $Y$ at $\beta = 0$,  $\hat{\gamma}$ can then be estimated as
$$
\hat{\gamma} = \frac{\mathrm{cov}\left(\lambda \Tilde{X} (Y-\mu), X(Y-\mu)  \right)}{\mathrm{var} \left( \lambda \Tilde{X} (Y-\mu) \right) } = \frac{\mathrm{var} \left( X\left(Y-\mu\right) \right)} {\lambda \mathrm{var}\left(\Tilde{X}(Y-\mu) \right) } = 1.
$$
The residual becomes 
$$r_i=X(Y-\mu) - \Tilde{X}(Y-\mu) = -U(Y-\mu),$$
Comparing $\mathrm{var}\left(r_i|\mathrm{stratum}\right)$ with ${\mathrm{var}\left(\mathbf{h}_i(0)|\mathrm{stratum} \right)}$, we end up with
$$
\mathrm{var}\left(r_i|\mathrm{stratum}\right)  = \frac{\mathrm{var} \left(U\left(Y-\mu\right)| \mathrm{stratum} \right)}{\mathrm{var} \left(X\left(Y - \mu \right) |\mathrm{stratum} \right)} {\mathrm{var}\left(\mathbf{h}_i(0)|\mathrm{stratum} \right)}
$$

\section{Case-control sampling for IPW estimator} \label{case-control}
Let us assume we have a rare disease ($E[Y]=p_0$) and modest covariate effects, so $p_i\ll 1$ for all $i$.  In the case stratum, the influence function $\mathbf{h}_i(\beta)=X_i(1-p_i)$, so
$$\mathrm{var}\left(\mathbf{h}_i(\beta)|Y=1\right)\approx \mathrm{var}\left(X_i|Y_i=1\right)\approx \mathrm{var}\left(X\right)$$
where the last approximate equality is exact if $\beta=0$.
In the control stratum, the influence function  $\mathbf{h}_i(\beta)=-X_ip_i$, so
$$\mathrm{var}\left(\mathbf{h}_i(\beta)|Y=0\right)\approx p_0^2\mathrm{var}\left(X\right).$$
This approximation is not as good as the case one, since the relative variation in $p_i$ is greater than that in $1-p_i$ for a rare disease: typically the control variance will be larger than $p_0^2\mathrm{var}(X)$. 
Neyman allocation says we need to take the population stratum sizes $N_h$ and the population stratum standard deviations $S_h$ and compute $N_hS_h$ for each stratum $h$. Under our approximations, these come to $Np_0\sqrt{\mathrm{var(X)}}$ for cases and $N(1-p_0)\sqrt{p_0^2\mathrm{var}(X)}$ for controls, which are approximately equal.  We should take the same number of cases and controls when covariate effects are small.

\bibliographystyle{plainnat}  

\newpage

\section*{Supplementary material}

\begin{figure}[h]
\centering
\caption{A comparison between the IF-IPW and IF-GR design of the first series of simulation studies}
\includegraphics[width=17cm]{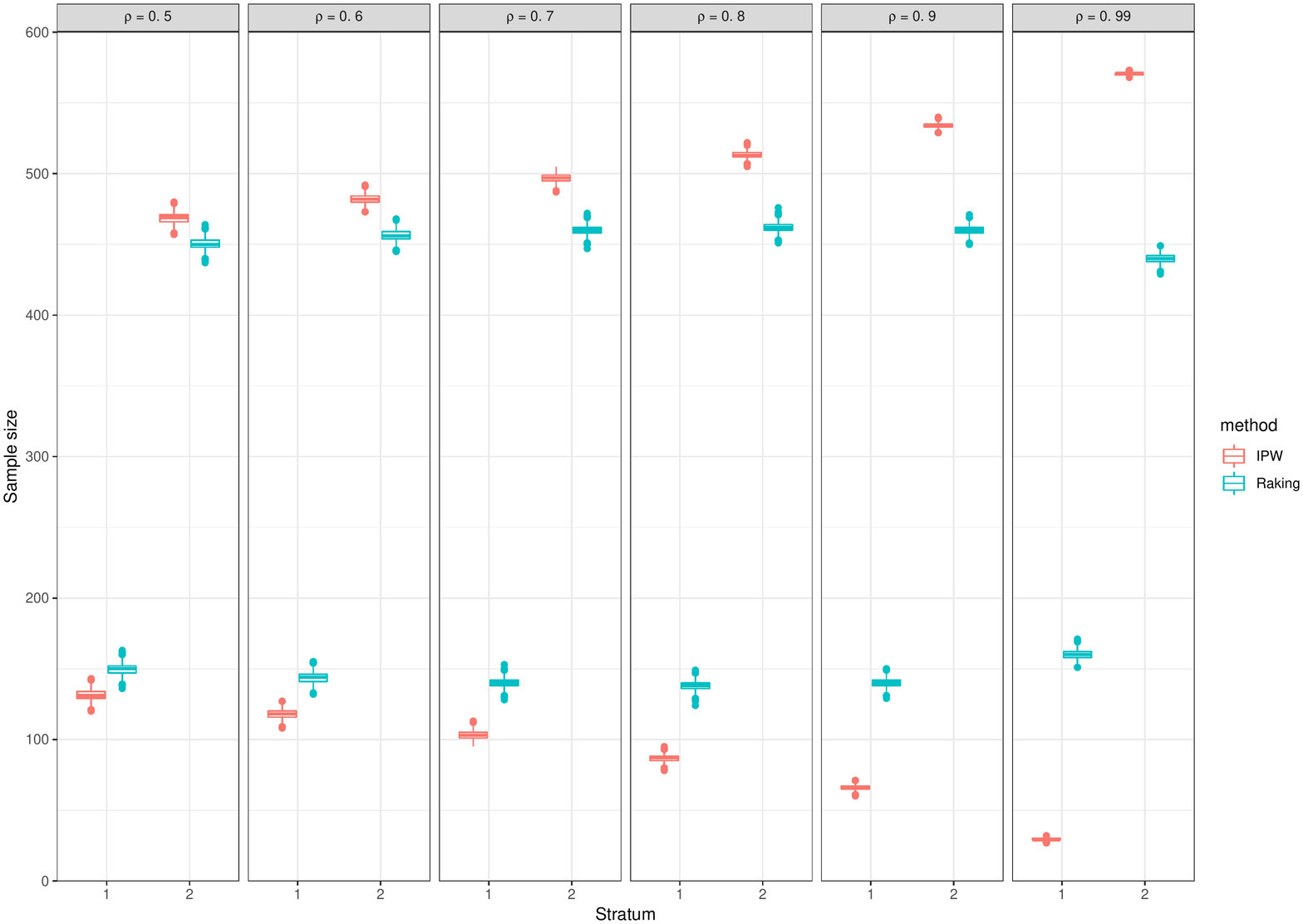}
\end{figure}

\begin{figure}[h]
\centering
\caption{A comparison between the IF-IPW and IF-GR design of the second series of simulation studies when $\sigma=0.5$}
\includegraphics[width=\textwidth]{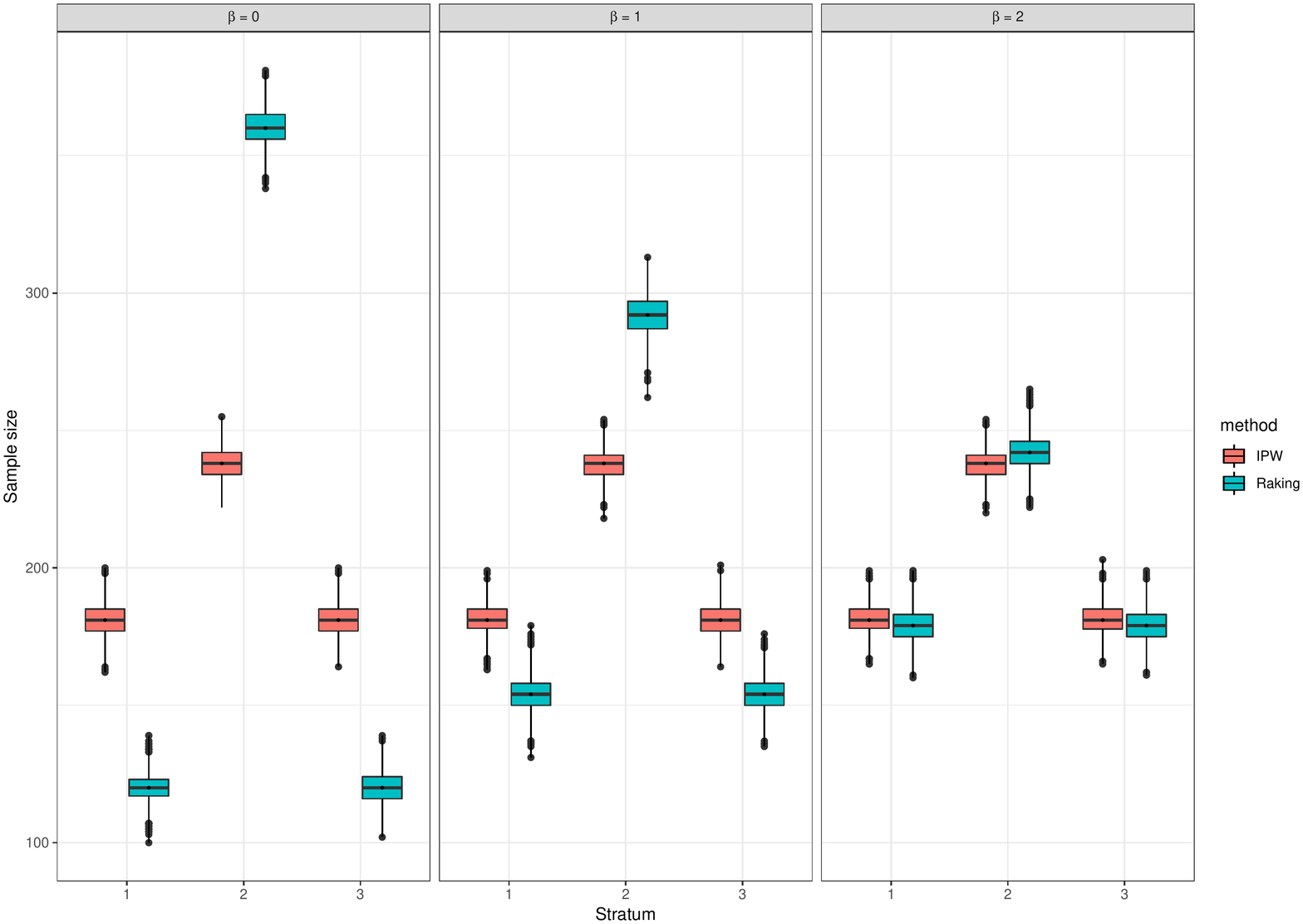}
\end{figure}

\begin{figure}[h]
\caption{A comparison between the IF-IPW and IF-GR design of the second series of simulation studies when $\sigma=0.75$}
\centering
\includegraphics[width=17cm]{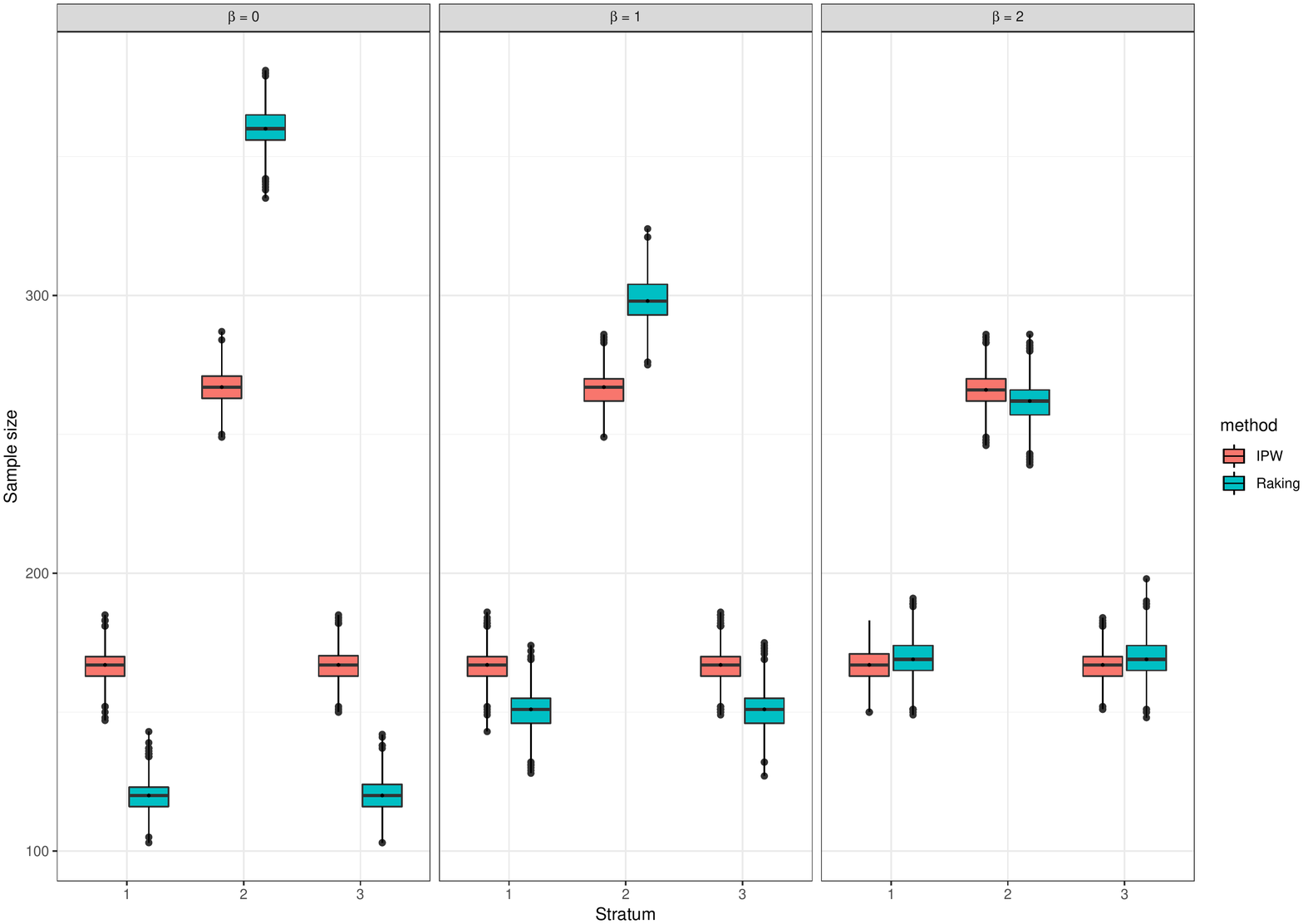}
\end{figure}

\begin{figure}[h]
\caption{A comparison between the IF-IPW and IF-GR design of the second series of simulation studies when $\sigma=1$}
\centering
\includegraphics[width=17cm]{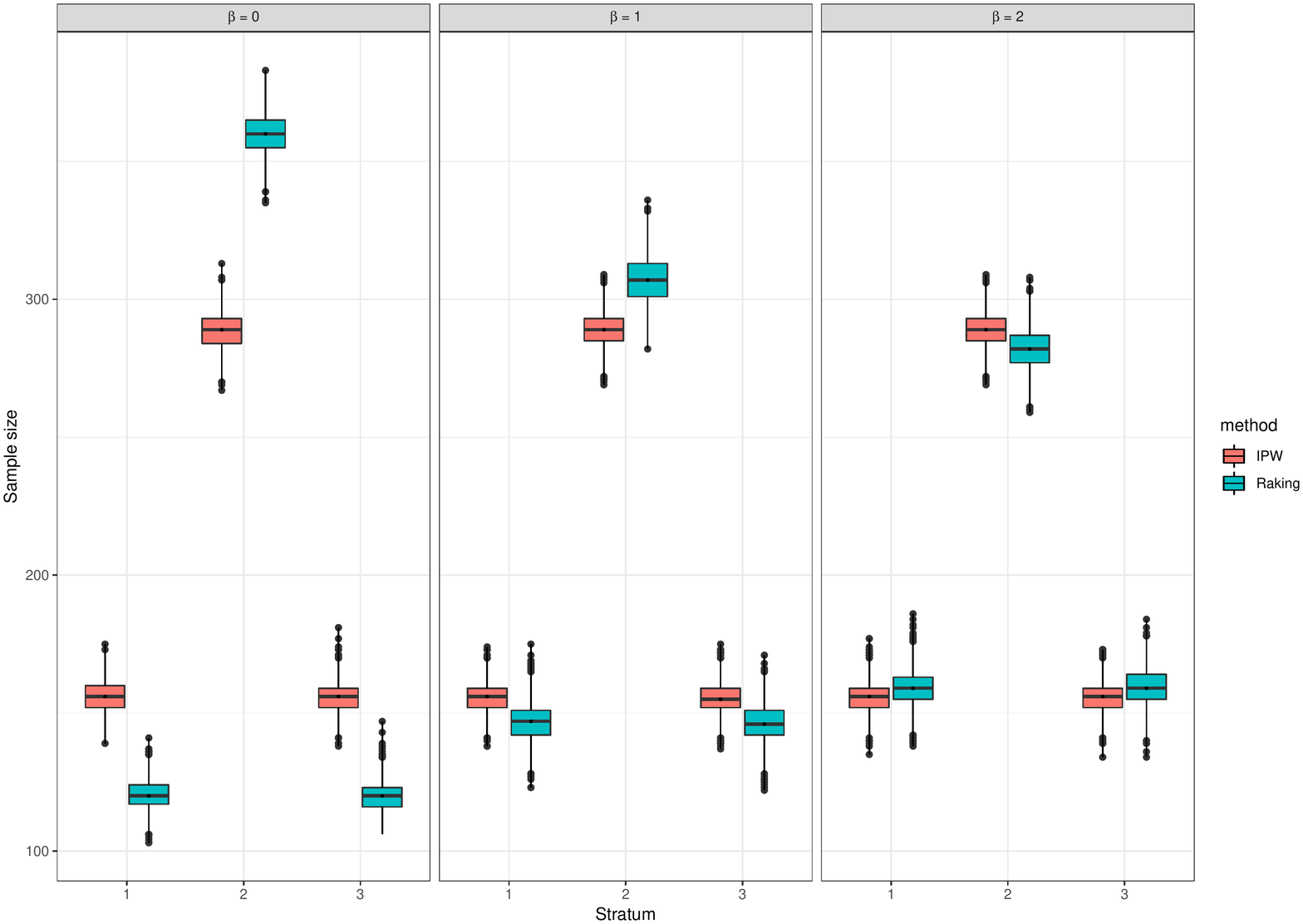}
\end{figure}

\begin{figure}[h]
\centering
\caption{A comparison between the IF-IPW and IF-GR design of the third series of simulation studies when $\sigma=0.5$}
\includegraphics[width=\textwidth]{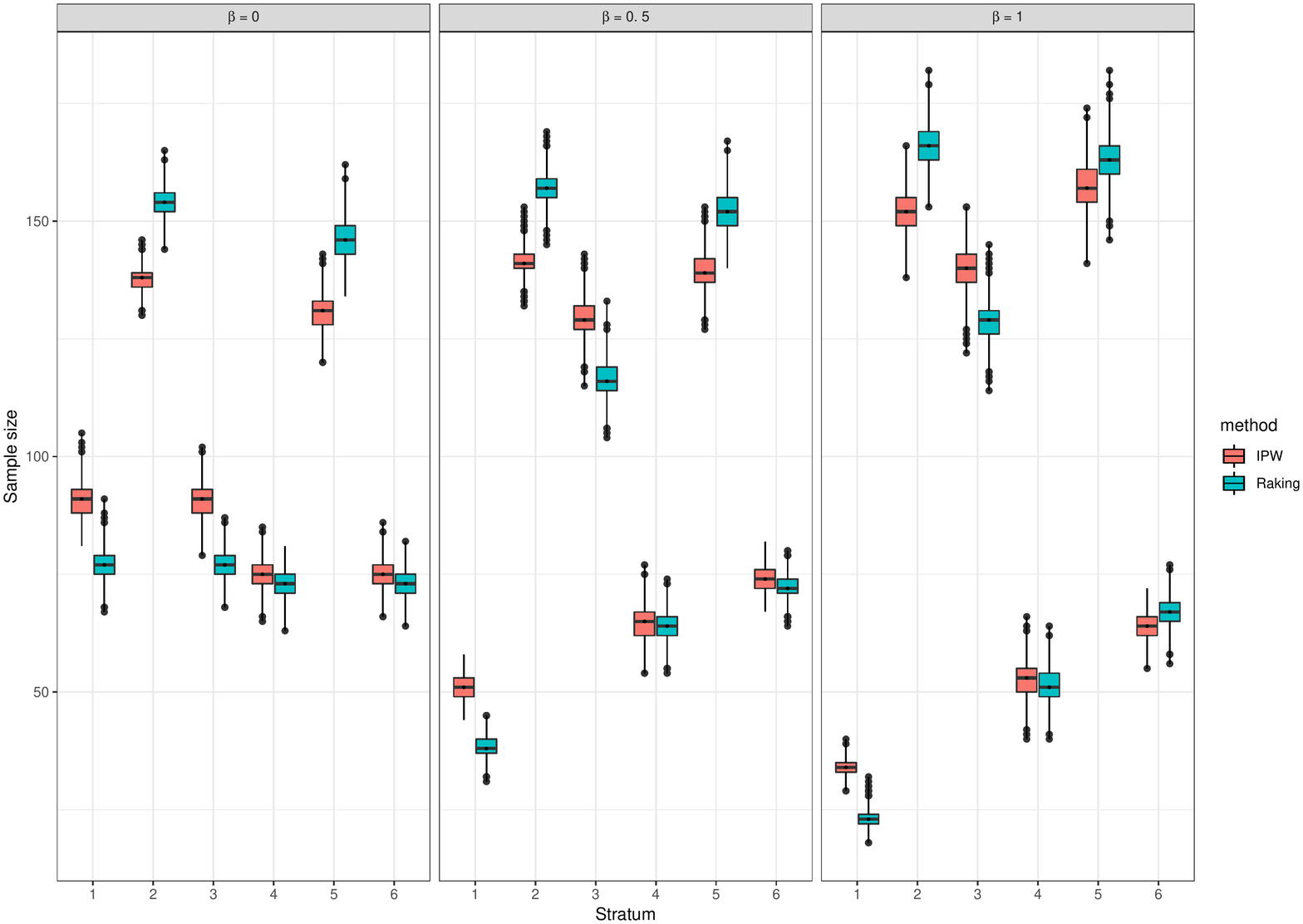}
\end{figure}

\newpage
\begin{figure}[h]
\centering
\caption{A comparison between the IF-IPW and IF-GR design of the third series of simulation studies when $\sigma=0.75$}
\includegraphics[width=17cm]{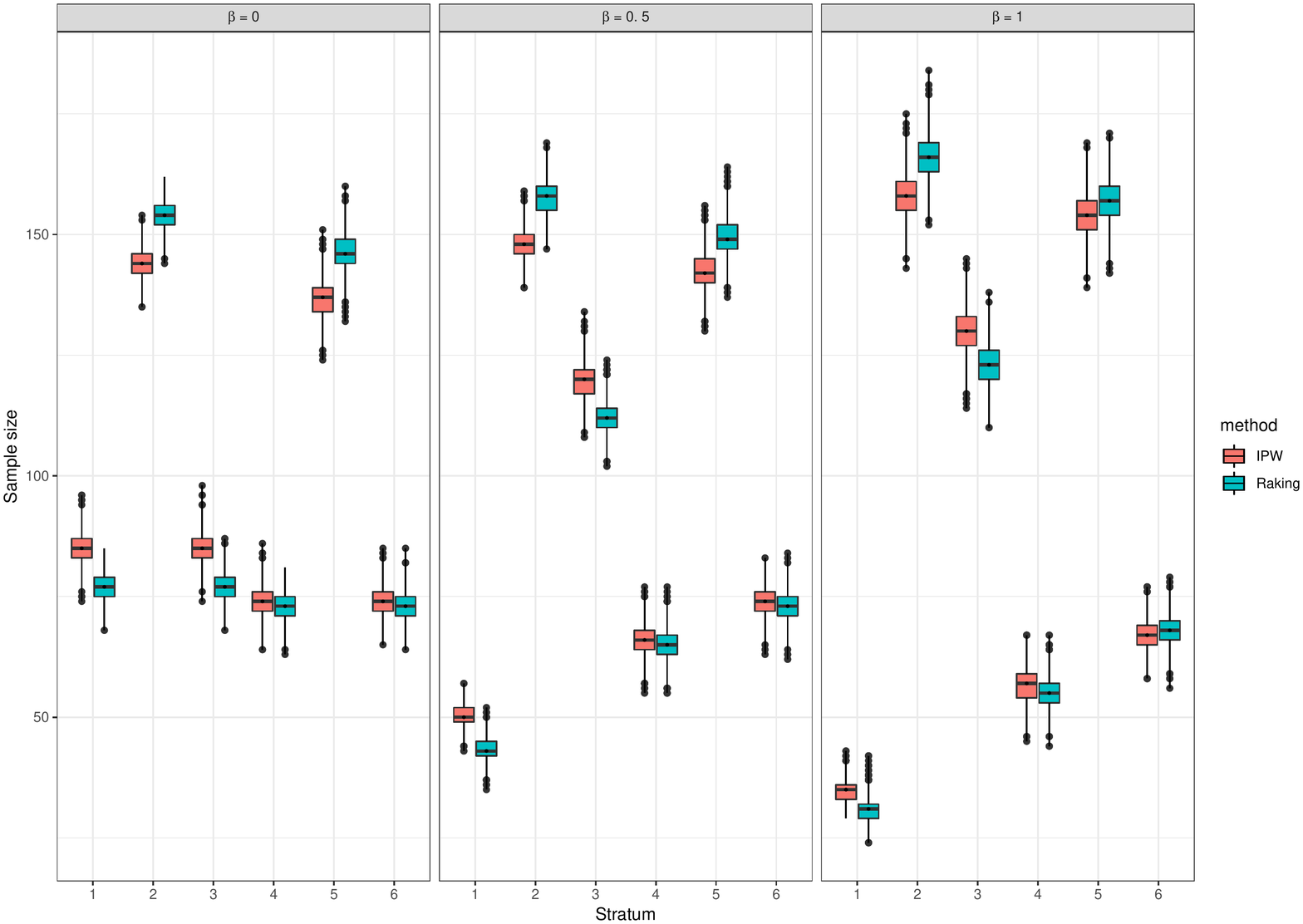}
\end{figure}

\begin{figure}[h]
\centering
\caption{A comparison between the IF-IPW and IF-GR design of the third series of simulation studies when $\sigma=1$}
\includegraphics[width=17cm]{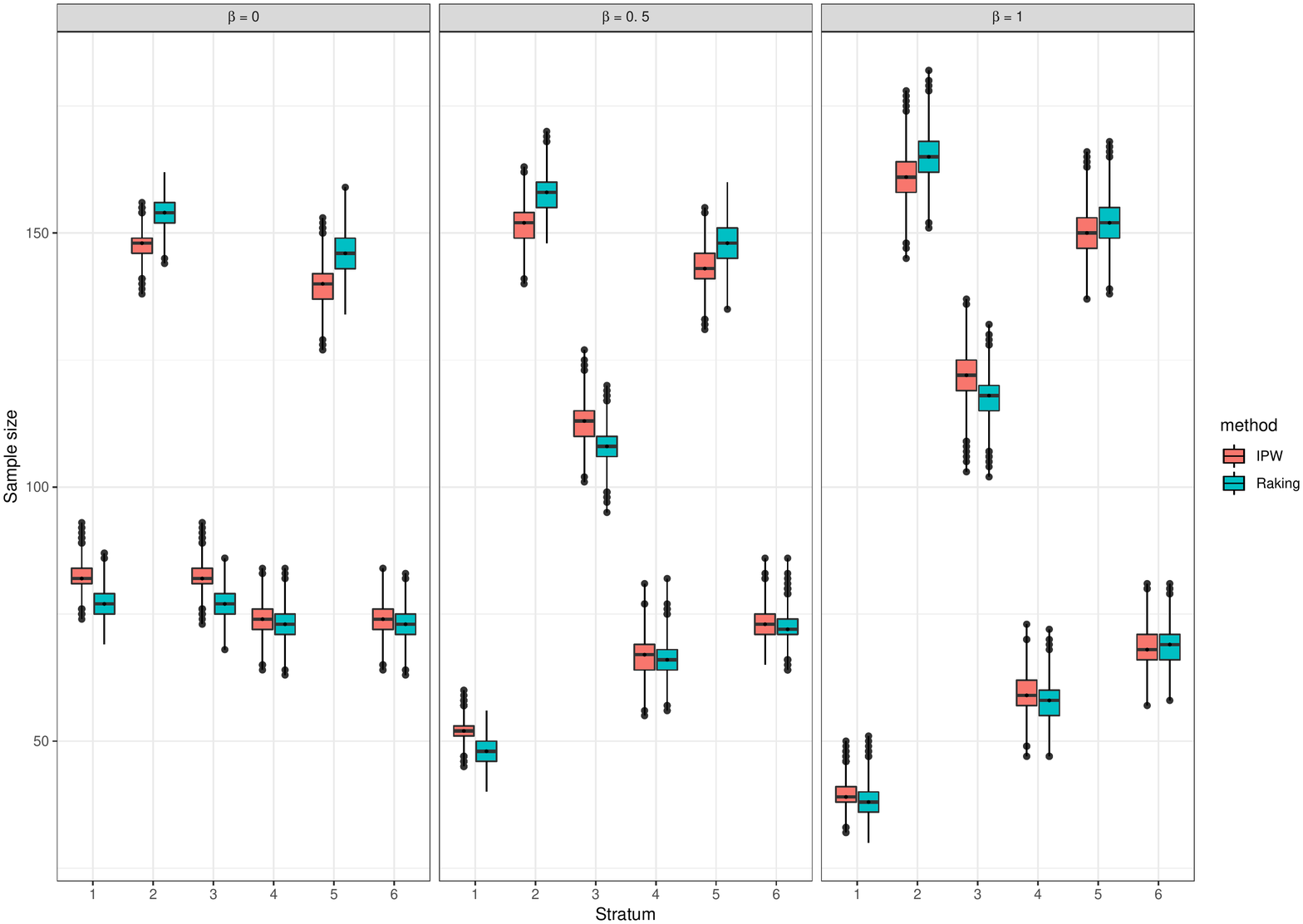}
\end{figure}

\begin{figure}[h]
\caption{The stratum-specific difference between the IF-IPW and IF-GR design of the last series of simulation studies when sensitivity and specificity are $0.95$}
\centering
\includegraphics[width=\textwidth]{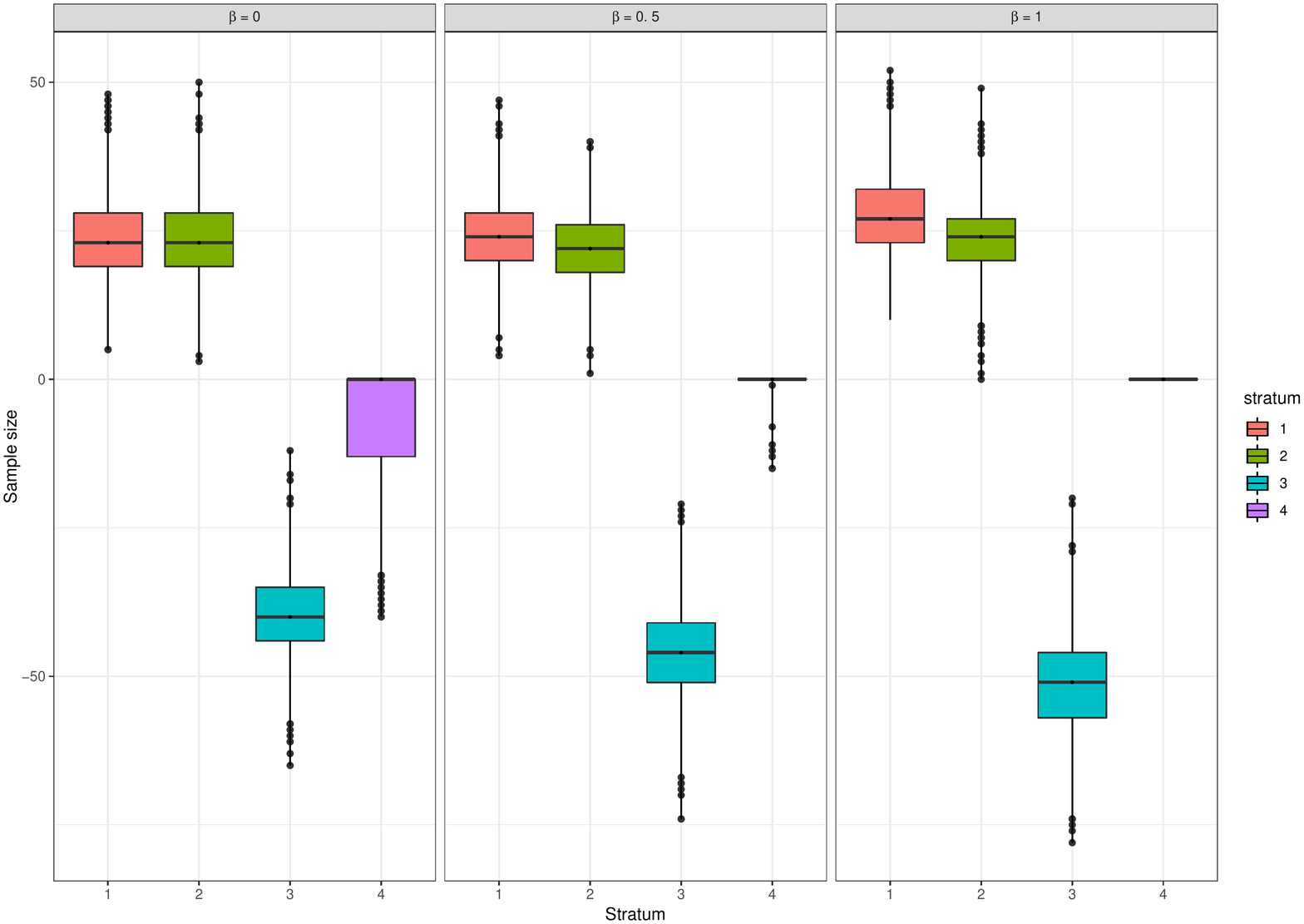}
\end{figure}

\begin{figure}[h]
\caption{The stratum-specific difference between the IF-IPW and IF-GR design of the last series of simulation studies when sensitivity and specificity are $0.9$}
\centering
\includegraphics[width=17cm]{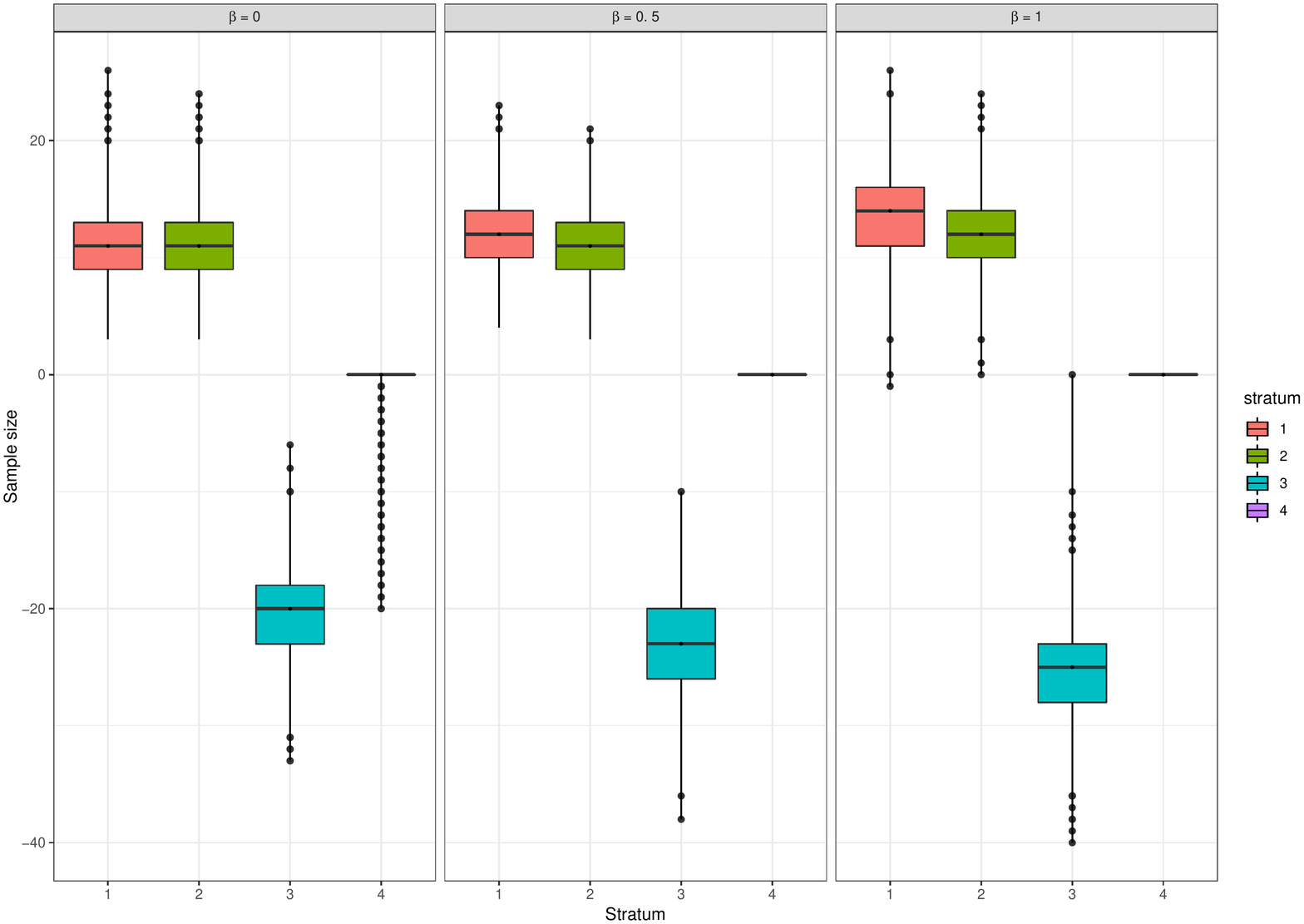}
\end{figure}

\begin{figure}[h]
\caption{The stratum-specific difference between the IF-IPW and IF-GR design of the last series of simulation studies when sensitivity and specificity are $0.85$}
\centering
\includegraphics[width=17cm]{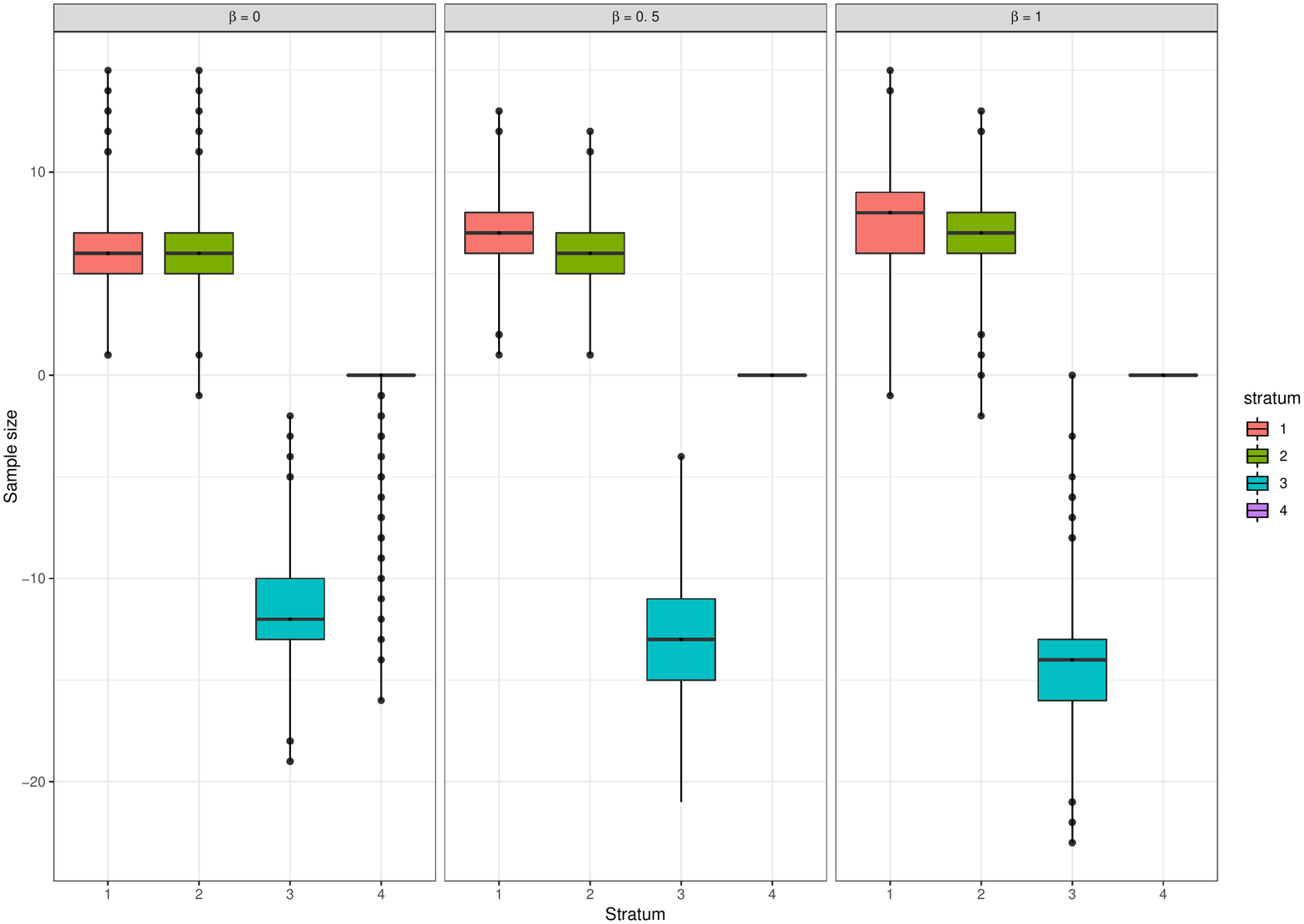}
\end{figure}

\end{document}